\documentstyle[12pt]{article}
\begin{document}
\begin{center}
{\Large \bf
Hadron spectroscopy based on relativistic Schr\"{o}dinger -
like wave equations
}\\
\vspace{0.4cm}
{\Large S.B.~Gerasimov} \\
\vspace*{.3cm}

{\small\it
Bogoliubov Laboratory of
Theoretical Physics,
Joint Institute for Nuclear Research, Dubna
}\\
\vspace{0.2cm}

{\large Abstract }
\end{center}

{\footnotesize
Relativistic potential-type equations are proposed
which approximately reproduce important features and constraints of
field-theoretical models and can be a useful tool in hadron spectroscopy.
Within this approach, the Regge-trajectory parameters
and some properties of higher radial excitations in the light quark sector
are discussed.
}.

\section{Introduction}

The phenomena in the long-distance (or the low-momentum) domain of
hadronic interactions are known to be
dominated by non-perturbative mechanisms of
QCD.  While the lattice QCD simulations are
widely accepted to provide the most direct
description of these interactions, they contain the inherent difficulties
connected with a finite lattice size and the violation of translational and
rotational invariance. Therefore, those nonperturbative methods
which maintain the mentioned symmetries are also needed, and the potential
model either in the nonrelativistic or relativistic wave equation context
belong to this category. Furthermore,
the continued accumulation of data on availability of hadrons
with the same quantum numbers, {\it e.g.}, the vector mesons
 produced in the $e^{+}e{-}$-annihilation, $\tau$-lepton decays,
photo- and electro-production reactions as well as continued and more
elaborated treatments of data \cite {BCS},\cite {D-K}
and phenomenological analyses of hadron
form factors \cite {DDW}  seem to favour the use of
more sophisticated mixing schemes of
constituent configurations in the considered resonance states.
So, the mixing of quarkonia $(q\bar q)$, the multiquark, {\it e.g.} $(q^2 {\bar
q}^2) $, and hybrid $(q\bar{q}g) $-states is observed as a really
important problem of hadron spectroscopy. We believe that
in solving this problem, those models will be advantageous which
deal with translation-invariant wave functions depending on a correct number
of degrees of freedom in the corresponding configuration space of two ( or
more) particles.  It seems natural to expect that the use of the same
approach based on 2-, 3-, or 4-body relativistic equations with effective
potential interactions between the constituents can present a sufficiently
consistent approximate scheme to consider the indicated problem before
resorting to more reliable but much more sophisticated
numerical approaches to the genuine nonperturbative lattice QCD.
In this
paper, we outline some features of the approach, where we are going to keep
as close as possible to known methods of dealing with one-particle
relativistic and few-body nonrelativistic wave equations while discussing the
spectroscopy of hadrons on the basis of Schr\"{o}dinger-like relativized
wave equations for bound quark-gluon systems.

\section{The orbital and radial trajectories of light mesons}

The valence quark model is known to be very successful in the
description of mesons and baryons as $q\bar q$- and $q^3$-states, where
quarks interact via effective potentials \cite {Lucha}.
The principally important
ingredient of these interactions is the linearly rising potential of
confinement. This result follows from the "quenched" approximation of
the lattice QCD giving the flavour-independent force between static colour
sources combined in the colour singlet states.
The "unquenching" procedure, {\it i.e.}, inclusion of
vacuum
polarization due to the light quark loops can result both in the system- and
state - dependence of effective potentials and in the modification
or even
termination of the linear behaviour of the confinement potential. A relevant
way to locate these effects is to study the properties of particle states
with the highest orbital and radial excitation of hadrons.

In this section, we report some results
on mass spectra and radial structure parameters of higher excitations of
quarkonia following the formulation proposed earlier \cite{Ger1} for the
potential approach to relativistic bound quark systems.
A specific feature of the formulated approach is
that the equations constructed include the squared forms of the Dirac
hamiltonians of each particle interacting with all other particles of a system
and additional variation conditions that define the one-particle energies via
the total energy of a system, the only spectral parameter entering into the
wave equation.

Here we concentrate on the topic of
Regge--trajectories for light quarkonia
to emphasize the special role of squared forms of
long-range potentials of confinement
in evaluation of characteristics of such states.
It is just the squared
form of the linear world-scalar potential of confinement that
 provides linearity of the Regge trajectories with the
 slope which can be made close to that following from the relativistic
string theory.  To illustrate this, we write first a general structure of
our relativistic two-body equation in the form \begin{equation}
\frac{1}{2W}(W^2 + \vec P^2 + M^2(p^2, r^2,..))\Psi (1,2)
=[\sum_{i=1,2} \frac{1}{2\varepsilon_i}(\varepsilon_i^2 + {\vec p_i}^2 + m_i^2) + \hat V] \Psi (1,2)
= W\Psi (1,2).
\end{equation}
\begin{eqnarray}
\vec P = \vec p_1 + \vec p_2, W = \varepsilon_1 + \varepsilon_2.
\nonumber
\end{eqnarray}
The solution of the eigenvalue problem, {\it i.e.} the finding of the
eigenmass $M$ for a given mass-operator $M^2$ in (1),
is seen to correspond to zero of the
inverse of the one-particle propagator written in the covariant form
\begin{eqnarray}
P_{\mu}P^{\mu} - M^2 = 0
\nonumber
\end{eqnarray}

The interaction kernel
$V$ in (1) is defined when in the equation for noninteracting
particles in their c.m.s.

\begin{equation}
W_0\Psi_0 (1,2) = [\sum_{i=1,2} \frac{1}{2\varepsilon_i}(\varepsilon_i^2 +\hat h_0^2(i)]\Psi_0 (1,2)
= [\sum_{i=1,2} \frac{1}{2\varepsilon_i}(\varepsilon_i^2 + {\vec p_i}^2
+ m_i^2)]\Psi_0 (1,2)
\end{equation}
we alternately replace $\hat h_{0}^2(i)$ by the squared forms of Dirac
operator $\hat h(i)$ for either particle in the field of another one
\begin{equation}
W\Psi (1,2) = \{[\omega_1 ( \frac{1}{2\varepsilon_1}(\varepsilon_1^2 +\hat h^2(1)) + \frac{1}{2\varepsilon_2}(\varepsilon_2^2 + \hat h_0^2(2))] + [1 \leftrightarrow 2]\}\Psi (1,2),\\
\end{equation}
\begin{equation}
\hat h^2(i) = \vec p_i^2 + m_i^2 + 2m_iV_s(r_{ij})
+V_s(r_{ij})^2 + 2\varepsilon_i V_v(r_{ij}) - {V_v}^2(r_{ij})
+ \mbox{spin-dependent terms},
\end{equation}

where
the weight factors $ \omega_i $ are normalized to unity:$ \omega_1+
\omega_2 = 1 $.If $ m_1=m_2 $, then, by symmetry arguments, one can expect
$\omega_1=\omega_2=\frac{1}{2}$.
When $m_1 \neq m_2$, we suggest to use the
simple relation $ \omega_1 / \omega_2 = \varepsilon_2 / \varepsilon_1$ that
reproduces, for the static limit $ m_j \to \infty  $, the correct form of the
corresponding one-particle equation in the field of a fixed center, {\it
e.g.}, the Klein-Gordon-Fock equation for spinless particles.
In our case,
the long-range confinement
"potential" is seen to be, in fact, the mass term parametrically
dependent on parton configuration coordinates.
As usual, in the spin-independent part of our
interaction kernel, we retain a world-scalar part and the 0-th
component of the vector interaction potential which survive in
nonrelativistic approximation.

To be phenomenologically acceptable, our relativistic equation with a given
Lorentz structure of the confining kernel should possess stable physical
solutions, {\it i.e.}, the binding energy should be real, and the state
should be localized in a finite region of the coordinate space. As the
presumed scalar and vector parts of the confining interaction squared enter
into the equation with opposite signs, the scalar part is required to be
stronger than the vector confinement potential: $ V_v \leq V_s$. The
squared Coulomb potential gives the most singular part of the interaction.
Hence for  typical scales of the ground state and higher radial states of
quarkonia with $l=0$, the effective $\alpha_{s}(r)$ values should be
bounded from above:
\begin{eqnarray} Im [\sqrt {(l + \frac{1}{2})^2 -
\frac{1}{2}(\frac{4}{3} \alpha_{s})^2}] = 0, \alpha_{s} \leq
\frac{3}{4\sqrt 2} \simeq .53 .
\end{eqnarray}

This means that the
effective coupling constant of QCD should be taken as "freezing" at the
value  $\sim .5$ by one of the proposed nonperturbative mechanisms (see,
e.g., \cite {Sol}) in the infrared region.

Below we are going to invoke also some ideas of the string-approximated
QCD, or rather, a string-like solution of the bag model \cite {J-T},
to further specify
the structure of the confining interaction and then to compare emerging
results with  the model-independent constraints \cite {Ji}
on general relativistic bound states.
It will be shown that the string-like structure of QCD, resulting
{\it e.g.}, from a semiclassical consideration of the rotating and
deformed ("fat-string-like") bag \cite {J-T} ,
can be represented by a much more simple potential model of
mesons that is able to approximately reproduce some important features and
results of the "microscopic" field-theoretical models.

The famous constant $B$ of the MIT-bag model is known to define the volume
energy of the space region inside an extended hadron, where nonperturbative
vacuum fluctuations of coloured fields are at least partially suppressed,
and it defines also the balance of the inward and outward pressure
of vacuum fields and the coloured fields created by partons of a given hadron.
This constant is considered to be connected with the contribution
to the total hadron mass of the "abnormal" part ({\it i.e.}, with the nonzero
trace) of the QCD energy-momentum tensor.
The general model-independent statement stressed by Ji \cite {Ji} is that
the ratios of contributions to the mass of the "abnormal"($\bar T_{\mu \nu}$)
and traceless ($\hat {T}_{\mu \nu}$) parts of the energy-momentum tensor are
\begin{eqnarray} \label{ratio}
\bar {M}/\hat {M} = 1/3,\\
\hat {M}/M_{tot} = 3/4.
\end{eqnarray}
The effective string-potential, or rather, the energy of the string-like
configuration of two static $\bar q q$-sources of a colour field, is found,
following \cite {J-T}, to be
\begin{eqnarray}
V_{s}(r) = a_{tot}r =(\bar{a} + \hat{a})r =(\frac {128}{3}\pi \alpha_{s}B)^
\frac {1}{2},
\end{eqnarray}
where $\bar{a} = \hat{a}$ represents, respectively, the contribution
of the chromoelectric field of quarks ({\it i.e.}, the traceless part of the
gluon energy-momentum tensor), and the bag energy is connected with the
gluon condensate, that is the trace-anomaly of QCD.

It was shown in \cite {J-T} that the classic consideration of rotation of
the chromoelectric "flux-tube" leads to a Regge-type relation between
the classic {\it nonquantized} angular momentum $J$  and mass $M_{string}$
\begin{eqnarray}
J = \frac {1}{2\pi a} M_{string}^2
\end{eqnarray}
where two parts of $M_{string}, \hat M_{string}$, and $ \bar M_{string}$
verify the general relations (6) and (7).

We use $V_{s}$ in the quantum-mechanical (quasi)potential two-body
equation treating it alternately as the effective mass of
one or another quark that parametrically depends on the interquark distance.
According to the given definition, we obtain for massless quarks, with
the neglect of spin-dependent terms,
\begin{eqnarray}\label{harm}
({\vec p}^2 + \frac{1}{2} a^2r^2 - \frac {1}{4} W^2)\Psi (1,2) = 0
\end{eqnarray}
The solution gives a simple dependence of the mass $M=M(l,n_r)$
on the {\it quantized} orbital ($l$) and radial ($n_r$) quantum numbers

\begin{eqnarray} \label{harmsolv}
W^2 \equiv M_{tot}^2 = 4\sqrt {2} a(l + 2n_r + \frac {3}{2}).
\end{eqnarray}
characteristic of a harmonic oscillator.
The harmonic quasipotential is obtained, however, after squaring
the linear world--scalar "potential" of confinement.
The adopted way of inclusion of the interaction kernel into our relativistic
two-body equation resembles the prescription of
the so-called "spectator"- type equation \cite{Gross}, where, alternately,
one of the particles is put off-mass-shell, while
the other is taken to be on-mass-shell.

According to (\ref {harmsolv}), the Regge-trajectory
slope $\alpha^{\prime}(0)=(4 \sqrt{2} a)^{-1}$ is within $10 \%$ of the
value $\alpha'(0)=(2 \pi a)^{-1}$ following from the relativistic string
theory.
Replacing formally $a$ in (\ref {harm}) by $\hat a = (1/2)a$, we keep in
$M^2$ only contributions of the traceless $\hat T_{\mu \nu}$-part of
massless quarks and gluons.  Therefore, we have
\begin{eqnarray}
\frac {\hat M^2}{M_{tot}^2} = \frac {1}{2}
\end{eqnarray}
which is again within $10 \%$ of the model-independent ratio (7).

The intercept $\alpha_{J}(0)$
of the leading $J=l+1$-trajectory is $-.5$ according to
(\ref {harmsolv}), hence unrealistic.
To calculate the realistic intercepts of Regge-trajectories,
one should include the spin-dependent potentials, presumably the
spin-orbit interaction induced by a scalar confinement "potential".

In this case, one can eliminate a still unknown parameter, using
the value of mass and known quantum numbers of a certain state
to evaluate masses of the states with other quantum numbers but with
the same quark content.
We note also that by analogy with the nonrelativistic case one can calculate
the mean value of the commutator of the mass-operator $M^2(p^2, r^2,..)$ with
the scalar product $(\vec p \vec r)$
\begin{eqnarray} \label {vir}
\langle \Psi_{l, n_r}| [M^2(p^2, r^2,..), (\vec p\vec r)]_{-}|\Psi_{l,
n_r}\rangle = 0
\end{eqnarray}
to get the "virial theorem", demonstrating that the contribution of
the mean value of the kinetic energy
of two massless valence quarks to the value of the total mass of a highly
excited meson is equal to the corresponding contribution of
the mean value of the "potential energy" that is connected with
the energy integrated over gluon degrees of freedom of a given hadron.
It is further tempting  to interpret this fact as giving a hint for
approximate equipartition, on the scale pertinent to a given bound state,
of the total momentum of a hadron in the "infinite momentum" frame between
the valence quark and gluon-sea partons,
the fact, following from the known moment of the nucleon structure function
measured on the scales of deep inelastic lepton-hadron scattering.

Further, we discuss further in brief some characteristics
of higher radial excitation of light vector quarkonia. This question
is especially timely in view of recent development of the Vector Meson
Dominance model applications \cite {DDW} to the analysis of nucleon
electromagnetic form factors both in the spacelike and timelike
regions of the transferred momenta $Q^2$. With the approximate equation
(\ref {harm}), we obtain "asymptotic" relations between masses of resonances
with high spins $J$, lying on the same trajectory, and masses of higher
radial excitations with the same spin
\begin{eqnarray}\label{radial}
m^2(J,n_r) - m^2(J^{'},n_r) = \alpha^{'}(0)^{-1}(J-J^{'})\\
m^2(J,n_r) - m^2(J,n_r^{'}) = 2\alpha^{'}(0)^{-1}(n_r-n_r^{'})
\end{eqnarray}
If we take $\rho(2130)$ \cite{PDG} as one of these higher radial states,
then the next two are $\rho(2600)$ and $\rho(3000)$
according to (15) and the value $\alpha^{'}(0) \simeq .9 GeV^{-2}$.
Curiously enough, a $\rho$ - type resonance with a
mass close to $2.6$ GeV was suggested in \cite{DDW} on the basis of
analysis of nucleon form factors.
The isoscalar partner
$\omega(3000)$ of the second state,
presumably, degenerated with it,
is seen to be near in mass to the $J/\psi$ - resonance, and it can play a
role in the enhancement of certain strong decays of $J/\psi$. This
situation deserves a more detailed study.

In the approach constructed, the energy functional has a recognizable
quasi-nonrelativistic form, therefore, the perspective is open up
to combine the accumulated experience in approximate solutions of the
spectral problems in the nonrelativistic (NR) domain with the description
of relativistic motion of hadron constituents.To this end, one should to
have preferably an analytic, although approximate, solution of the
corresponding NR problem. For example, in our case, one can simply estimate
the dependence of the "psi-at-zero" values

\begin{eqnarray}\label{psi-nr}
\psi (0) = \frac{m_{red}}{2\pi}<V^{\prime}(r)>
\end{eqnarray}

on masses or
quantum numbers of corresponding meson states.  Making use of the analogy
of our equation (\ref {harm}) with the nonrelativistic Schr\"{o}dinger
equation, we obtain approximate scaling relations for the "psi-at-zero" in
the case of highly relativistic radially-excited states, hence, for
leptonic widths of the corresponding resonances

\begin{eqnarray}\label{psi-rel}
\psi (0)^2 = \frac{a^2}{4\pi}<r> = \frac{am_{V_{n}}}{\sqrt{6\pi ^3}}\\
\Gamma_{ee}(V_n) \sim \psi (0)^2/m_{V_n}^2 \sim (n-1/4)^{-\frac{1}{2}}
\end{eqnarray}

where $m_{V_n}$
is the mass of the radially excited $nS$-state of the vector resonance;
$n=n_r+1$ , the principal quantum number; $n_r$,  the radial quantum
number.  To get the absolute values of these leptonic widths, one should
consistently include the QCD radiative corrections into the amplitude of
 $\bar q q \rightarrow e^{+}e^{-}$ transition.  The estimation of total
widths of higher radial excitations of vector mesons can proceed as
follows. With the meson mass of order $2~GeV$ or larger, many decay
channels are open, and one can rely on the quark-hadron duality idea while
assuming the dominant role of the initial quark-parton stage of the
reaction that defines the width dependence on quantum numbers and mass of
the resonance. The total $q\bar q$ cross-section in the $S$-state is
assumed to scale as $m_{V_n}^{-2}$, where $m_{V_n}$ is the mass of the
$V_{n_{r}}$-meson. At the hadronization stage, we adopt the simplest
phase-space correction, assuming its form from a presumably main
decay mode. For the isovector $\rho_{n}$-type resonances, the
apparently conspicuous or, at least, important decay channel is the
$\rho(.77~GeV)\pi\pi$-channel. With adopted assumptions, one can get
\begin{eqnarray}
\Gamma_{tot}(V_{n+1}) \simeq \Gamma_{tot}(V_{n}) \frac
{m_{V_{n+1}}f(\alpha_{n+1})} {m_{V_{n}}f(\alpha_{n})}\\
f(\alpha_{\rho_{n}}) = 1-\alpha^4+4\alpha^2log\alpha
\end{eqnarray}

where $\alpha_{n} = m_{\rho}/m_{\rho_{n}}$, and the phase-space formula for
the decay $\rho_{n}(m_{n}) \rightarrow \rho(.77) 2\pi$ has been obtained
\cite{Kop} for massless pions. So, with the input
$(m=2.6~GeV;\Gamma=.6~GeV)$ for mass and width of the heaviest claimed
$\rho$-type resonance, we get for the next two resonances lying below
the open charm threshold:  $(mass;width) \rightarrow (2.98;.77)$ and
$(3.34;.92)$.  The very large and, seemingly, overestimated  widths
suggest, nevertheless, that hadronic corrections can be important in
calculations of masses of high radially-excited states. This problem is
still to be considered.

Concerning perspectives of the study of higher excited states
within the outlined approach, one can notice that the lattice QCD
simulations with the unquenched light quarks seem to be consistent with
the linear behaviour of the confinement potential up to the distance
of $r\le 2~fm$ \cite{Aoki}. With the help of the virial theorem (2.13)
one can relate the mean distance $\sqrt{r^{2}_{q\bar q}} \le 2~fm$ between
the $q\bar q$-pair with the corresponding quantum numbers ($l\le 15;
n_{r}=0$), or ($n_{r}\le 9;l=0$) of a given bound state which are
considerably larger than the corresponding values $l\simeq 6\div 7$ or
$n_{r}\simeq 4\div 5$ of the known meson resonances.  Hence,
it seems that many
resonances could still be explored theoretically on the basis of
relativistic wave equations with the linear confinement term, playing the
role of part of effective quark (or gluon) mass, parametrically
dependent on the distance between corresponding force centers.

\section*{Acknowledgments}
The author is grateful to the Organizing Committee of the
NUPPAC'99 Conference, 13-17 November 1999, Cairo, Egypt, for the invitation
and partial support.


\begin{thebibliography}{99}

\bibitem{BCS}
T. Barnes, F.E. Close and E.S. Swanson, {\em Phys.Rev.\/} {\bf D 52} (1995)
5242.

\bibitem{D-K}
A. Donnachie and Yu.S. Kalashnikova, {\em E-print Archive:
hep-ph/9901334}.

\bibitem{DDW}
S. Dubnicka, A.Z. Dubnickov\'a and P. Weisenpacher, in {\em
Hadron Structure '98}, Proc. of the Int. Conference, Stara Lesna,
Slovak Republic, 7-13 Sept. 1998, D. Bruncko and P. Strizenec, eds., IEP SAS,
Kosice, p.127.

\bibitem{Lucha} W.Lucha, F.F.Sch\"{o}berl and D.Gromes,
{\em Phys. Rep.\/} {\bf 200} (1991) 127.

\bibitem{Ger1}
S.B.Gerasimov, in
{\em Prog.Part.Nucl.Phys.\/} {\bf 8}, Sir D. Wilkinson, ed., Pergamon Press,
1982, p.207;

{\em Multiquark interactions and QCD\/}, JINR-D1,2-81-728, Dubna, 1981, p.51;

{\em Special Topics in Gauge Field Theory\/},
PHE-82-10, Berlin-Zeuthen, 1982, p.31.

\bibitem{Sol}
A.C. Mattingly and P.M. Stivenson, {\em Phys.Rev.\/} {\bf D 49} (1994) 437.

Yu.A. Simonov, {\em Yad.Fiz.\/} {\bf 58} (1995) 1139.

D.V. Shirkov and I.L. Solovtsov, {\em Phys.Rev.Lett.\/} {\bf 79} (1997) 1209.

\bibitem{J-T}
K. Johnson and C.B. Thorn, {\em Phys.Rev.\/} {\bf D 13} (1976) 1934.

\bibitem{Ji}
X. Ji, {\em Phys.Rev.Lett.\/} {\bf 74} (1995) 1071;
{\em Phys.Rev.\/} {\bf D 52} (1995) 271.

\bibitem{Gross} F.Gross, et.al., {\em Phys. Rev.\/} {\bf C 45} (1992) 2094.

\bibitem{PDG} Review of Particle Properties, {\em Phys. Rev.\/} {\bf D 50}
(1994).

\bibitem{Kop} G.I.~Kopylov, {\it Osnovy kinematiki resonansov}, M.
"Nauka", 1970, s.246$\div$247.

\bibitem{Aoki} S.~Aoki, et al. (CP-PACS Collaboration), {\it E-print
Archive: hep-lat/9902018}.

\end{thebibliography}
\end{document}